\documentclass[aps,pre,twocolumn,showpacs,10pt]{revtex4-1}
\usepackage{graphicx}
\usepackage{epstopdf}
\usepackage{amsmath}
\usepackage{amssymb}
\usepackage{default}
\usepackage[english]{babel}
\newcommand{\be}{\begin{equation}}
\newcommand{\bea}{\begin{eqnarray}}
\newcommand{\bc}{\begin{center}}            
\newcommand{\ee}{\end{equation}}
\newcommand{\eea}{\end{eqnarray}}
\newcommand{\ec}{\end{center}}

\newcommand{\baa}{\begin{eqnarray*}}
\newcommand{\eaa}{\end{eqnarray*}}
\begin{document}
\title{Constraints on Onsager coefficients from quasi-static and reversible
operations}
\author{Ramandeep S. Johal}
\email{rsjohal@iisermohali.ac.in}
\affiliation{Department of Physical Sciences, \\ 
Indian Institute of Science Education and Research Mohali,\\  
Sector 81, S.A.S. Nagar, Manauli PO 140306, Punjab, India}
\begin{abstract}
The performance of a generic, cyclic heat engine between two heat reservoirs is
discussed within a linear-irreversible framework. The 
Onsager reciprocal relation is derived
as a consequence of the equivalence between quasi-static and reversible
operations, under the tight-coupling condition. When the latter condition is 
relaxed, it is possible to achieve reversible cycle 
in a finite duration. Onsager reciprocity must be violated
when either the quasi-static cycle is not reversible, or the reversible
cycle is not quasi-static.
\end{abstract}
\maketitle
A quasi-static process, despite being an idealized concept, is
of great importance in equilibrium thermodynamics \cite{Callenbook1985}.
Several textbook models of heat cycles
such as Carnot cycle, Otto cycle and so on, are based on these processes
which run infinitely slowly, but may or may not be reversible. Further, 
any framework for nonequilibrium processes involving
rates and fluxes, quite understandably, must 
approach the limiting case of equilibrium theory when these fluxes become
negligible in response to vanishing thermodynamic forces.
In particular, in the linear response regime \cite{Onsager1931a}, the generalized fluxes are
assumed to be linear functions of the forces. The resulting 
linear-irreversible framework is immensely successful
in unifying diverse, coupled phenomena such as Seebeck, Peltier, Dufour effects and so on
\cite{Callen1948, Joubook}. 

In this letter, we apply the linear-irreversible framework to  
a heat engine operating between two heat reservoirs. The restriction
to near-equilibrium conditions demands that
the temperature difference between the reservoirs is small and 
the duration of the heat cycle ($\tau$) is sufficiently long.
Further, the generalized fluxes are not defined as instantaneous quantities
(time derivatives of macroscopic variables, as in Onsager formalism \cite{Onsager1931a}), 
but as time-averages over one cycle ($X/\tau$).
We determine the constraints on the  
Onsager coefficients in the flux-force relations, 
taking into account the ideal limits of quasi-static and 
reversible cycles. A connection will be established 
between the reciprocal property of Onsager coefficients 
and the equivalence of quasi-static and reversible cycles \cite{commentqsrev}.
It is shown that violation of this reciprocity implies that 
quasi-static and reversible cycles are not 
equivalent. Further, it is possible to construct a 
reversible cycle in a finite duration, or in other words,
achieve a finite power output at Carnot efficiency.

Let $Q_h$ denote the heat absorbed in one cycle from the hot 
reservoir at temperature $T_h$, and $Q_c$ be  
the amount of heat rejected to the cold reservoir at temperature $T_c$.
The work performed by the engine is
\be
W = Q_h -Q_c > 0,
\label{w}
\ee
while the total change in the entropy of the reservoirs is
\be 
\Delta  S = \frac{Q_c}{T_c} -\frac{Q_h}{T_h}.
\label{dstot}
\ee 
The working medium undergoes a cycle and so does not involve 
a net change in its entropy. 
Now, if $Q_c$ is treated as a floating variable, it may be
eliminated from Eqs. (\ref{w}) and (\ref{dstot}), 
so that the second law inequality ($\Delta  S \geq 0$)
implies $W \leq Q_h \eta_{\rm C}^{}$,
where $\eta_{\rm C}^{} = 1-T_c/T_h$ is the Carnot efficiency.
Thus, the second law sets an upper bound for work as $W_{\rm rev} = Q_h \eta_{\rm C}^{}$, obtained 
under a reversible operation ($\Delta  S = 0$).
In general, we can write
\be
W = W_{\rm rev} - T_c \Delta  S, 
\label{gs}
\ee 
The deficit, $W_{\rm rev}-W$, is referred to as
the lost work---the energy that is not available for work during the
irreversible operation \cite{Zemansky1957}.  

Now, let us consider the time-dependence of the energy-conversion process. 
The rate of total entropy production per cycle, $\dot{S} \equiv {\Delta  S}/{\tau}$,
can be written as:
\be
\dot{S} = \frac{1}{\tau} \left(\!-\frac{W}{T_c} \!\right) + 
\dot{Q}_h \left( \frac{1}{T_c} - \frac{1}{T_h} \right), 
\label{dots}
\ee 
where $\dot{Q}_h = Q_h/\tau$. Following
 the flux-force framework of linear-irreversible thermodynamics,
we identify two generalized fluxes ($J_i$) and the
corresponding thermodynamic forces ($X_i$):
\bea
{J}_1 &=& \frac{1}{\tau}, \qquad X_1 = -\frac{W}{T_c},
\label{j1ld}\\
{J}_2 &=& \dot{Q}_h, \qquad X_2 = \frac{1}{T_c} - \frac{1}{T_h},
\label{j2ld}
\eea
which imply $\dot{S} \equiv {J}_1 X_1 + {J}_2 X_2$.
$J_1$ is the number of cycles executed per unit time, and so
represents the rate at which the heat cycle proceeds.
Since the forces and the fluxes are assumed to be small,
we are dealing here with long cycle durations close to
equilibrium.
The process 2, satisfying $J_2 X_2 >0$, denotes a spontaneous process
leading to positive entropy production, while the process 1
with $J_1 X_1 <0$ is the driven or non-spontaneous process which 
cannot proceed in the absence of process 2.
The linear regime implies that the flux-force relations are
in the form: $J_i = \sum_{j=1}^{2} L_{ij} X_j$, where $i=1,2$.
Here, the phenomenological coefficients $L_{ij}$ will be referred
to as Onsager coefficients. They are assumed  fixed
within the regime of small forces. Then, $\dot{S}$ becomes   
a binary quadratic form in $X_1$ and $X_2$, and the inequality 
$\dot{S} \geq 0$ imposes the following conditions:
\be
L_{11}, L_{22} \geq 0, \quad
%
4 L_{11} L_{22} \geq  \left( L_{12} + L_{21} \right)^2.
\label{Lcond}
\ee
There seems no reason, per se, to assume  
the reciprocal relation ($L_{12} = L_{21}$) which was 
originally derived by Onsager from the principle of microscopic 
reversibility and the theory of equilibrium fluctuations \cite{Onsager1931b}. 

Explicitly, the flux-force relations are given by: 
\bea
\frac{1}{\tau}  &=& -{L}_{11} \frac{W}{T_c} + {L}_{12} \frac{\eta_{\rm C}^{}}{T_c},
\label{j1glo} \\
\frac{Q_h}{\tau} &=& -{L}_{21}\frac{W}{T_c}  + {L}_{22} \frac{\eta_{\rm C}^{}}{T_c}.
\label{j2glo}
\eea
Assuming $L_{11} >0$, we can eliminate $W$ from the above pair of equations, to write:
\be
\frac{Q_h}{\tau} = \frac{L_{21}}{L_{11}} \frac{1}{\tau} + 
\frac{\vert {\bf L} \vert}{L_{11}} \frac{\eta_{\rm C}^{}}{T_c},
\label{ff12}
\ee
where the determinant $\vert {\bf L} \vert \equiv 
{L_{11}}{L_{22}} - {L_{12}}{L_{21}}$ satisfies $\vert {\bf L} \vert \geq 0$.
Let us first assume, $\vert {\bf L} \vert = 0$,
which is also known as the tight-coupling condition \cite{Caplan1965}.
Under this condition, 
the fluxes $J_1$ and $J_2$ become proportional to each other. 
Therefore, Eq. (\ref{ff12}) yields:
\be
Q_h = \frac{L_{21}}{L_{11}} >0,
\label{qh2}
\ee
which requires $L_{21} > 0$. Then, from the $\vert {\bf L} \vert = 0$ condition, 
we have $L_{12} \geq 0$. 

Now, consider the approach towards the quasi-static operation ($\tau \to \infty$)
in Eq. (\ref{j1glo}). The work output in this limit
is given by: $W_{\rm qs} =(L_{12}/L_{11}) \eta_{\rm C}^{}$. Then,
Eq. (\ref{j1glo}) may be rewritten in the form:
\be
W = W_{\rm qs} - \frac{T_c}{L_{11}\tau}.
\label{j1w}
\ee
Thus, the work output may be 
controlled by varying $\tau$, 
while the magnitude of $Q_h$ is kept fixed. 
So, as $\tau$ is decreased, the work output 
decreases and vanishes at the minimum duration: 
$\tau_{\rm min} = {T_c}/{L_{12} \eta_{\rm C}^{}}$.

It is well known that a 
quasi-static operation may or may not imply a reversible
operation \cite{Callenbook1985, commentcycles}. 
If it does, then as $\tau \to \infty$ for our cycle, we also have 
$\Delta  S \to 0$. This equivalence has 
a number of interesting consequences.
We can now write: 
$W_{\rm qs} = W_{\rm rev}$, which yields  $Q_h = {L_{12}}/{L_{11}}$. 
Upon comparison with Eq. (\ref{qh2}), we obtain:
\be
L_{12} = L_{21},
\label{reci}
\ee
thus obtaining the reciprocal relation for Onsager coefficients.  
Conversely, from the general relations derived above, 
we can write: $W_{\rm qs} = (L_{12}/L_{21}) W_{\rm rev}$.
Thus, the validity of Onsager reciprocity for macroscopic
linear-irreversible, cyclic engines implies that quasi-static
and reversible operations of the engine are equivalent. 
This is the first main result of this letter.

Furthermore, under the condition  $W_{\rm qs} = W_{\rm rev}$,
the comparison of Eqs. (\ref{gs}) and (\ref{j1w}) yields:
\be
\Delta  S = \frac{1}{L_{11}\tau}.
\label{sld}
\ee
Thus, the total entropy produced
varies inversely with the cycle duration---consistent with 
the assumption that the quasi-static
operation implies reversible operation.

On the other hand, a quasi-static process may not be
reversible, and involve 
some entropy production $\Delta  S_{\rm qs} >0$. 
A familiar situation is the presence of friction
between different parts of the engine, which may
not vanish even if the cycle is made infinitely
slow. Then, from Eq. (\ref{gs}), we can write: 
$W_{\rm qs} = W_{\rm rev} - T_c \Delta  S_{\rm qs}$.
So, in general, we have: $W_{\rm qs} \leq W_{\rm rev}$,
which implies that $L_{12} \leq L_{21}$.
Thus, the condition $L_{12} < L_{21}$ represents the 
fact that the quasi-static work is less than the reversible
work---indicating the presence of friction or viscous forces.
Related to this, we also have the result for quasi-static
efficiency:
$\eta_{\rm qs} = W_{\rm qs}/Q_h = 
(L_{12}/L_{21})  \eta_{\rm C}^{} \leq \eta_{\rm C}^{}$.
\par
If $\vert {\bf L} \vert >0$, then from Eq. (\ref{ff12}), $Q_h$
diverges in the quasi-static limit \cite{commentleaks}. Since $W_{\rm qs}$ is 
finite, so the efficiency vanishes in the 
quasi-static limit. Clearly, the latter does not imply 
a reversible cycle in the $\vert {\bf L} \vert >0$ case. 
A natural and intriguing question here is whether it is possible to 
achieve reversible operation   
in a finite duration (since $Q_h$ is finite). 
In recent years, the possibility of 
achieving the Carnot efficiency in finite time, or
in other words, a finite power output along with Carnot efficiency, 
has attracted attention \cite{Benenti2011, Armen2013, Brandner2015, Polettini2015,
Proesman2015, Campisi2016}. 
We now approach this issue for the case of cyclic, linear-irreversible engines.

Eliminating $\tau$ from Eqs. (\ref{j1glo}) and 
(\ref{j2glo}), and using the reversible work condition, $W = Q_h \eta_{\rm C}^{}$,
we obtain the quadratic equation:
$L_{11} Q_{h}^{2} - (L_{12}+L_{21})Q_h + L_{22} = 0$,
whose solutions are:
\be
Q_h = \frac{L_{12}+L_{21} \pm \sqrt{(L_{12}+L_{21})^2 - 4 L_{11} L_{22}}}{2 L_{11}}.
\label{quadroot}
\ee
The only real solution in the above is obtained when the 
the third condition in Eq. (\ref{Lcond}) reduces to an equality:
\be
(L_{12}+L_{21})^2  = 4 L_{11} L_{22},
\label{3rd}
\ee
and therefore
\be
Q_h = \frac{L_{12}+L_{21}}{2 L_{11}} \geq 0.
\label{qhroot}
\ee
The magnitude of the reversible work---performed in a finite duration---is then given by
$W_{\rm rev}^{(\tau)} = (L_{12} + L_{21}) \eta_{\rm C}^{} / 2 L_{11}$.
Note that the above expression for $Q_h$ holds specifically 
for the reversible operation. 
Unlike the tight-coupling case ($\vert {\bf L} \vert =0$) 
that yields a fixed magnitude for $Q_h$ (Eq. (\ref{qh2})),
here we have $\vert {\bf L} \vert \neq 0$, and so 
in general, $Q_h$ depends on the cycle duration (Eq. (\ref{ff12})).

The duration $\tau_{\rm rev}$ of the reversible cycle can be calculated 
from Eqs. (\ref{j1glo}) and (\ref{qhroot}) as:
\be 
\cfrac{1}{\tau_{\rm rev}} = ({L_{12} - L_{21}})\cfrac{\eta_{\rm C}^{}}{2T_c},
\label{taurev}
\ee
which requires $L_{12} \geq L_{21}$. Using the positivity condition on the 
quasi-static work, $W_{\rm qs} =(L_{12}/L_{11}) \eta_{\rm C}^{} >0$, 
we can set $L_{12} >0$. 
Then, the inequality $L_{12}+L_{21} \geq 0$ in Eq. (\ref{qhroot})
implies that $L_{21} \geq - L_{12}$.
These considerations constrain 
the possible values of $L_{21}$ as follows:
\be
-L_{12} \leq L_{21} \leq L_{12}.
\label{1221}
\ee
Since, the minimum allowed value of $L_{21}$ is $-L_{12}$, so the minimum value of 
${\tau_{\rm rev}}$ is ${T_c}/{L_{12} \eta_{\rm C}^{}} \equiv {\tau_{\rm min}}$
at which the work output vanishes. 
On the other hand, for $(L_{12} -L_{21}) \to 0^+$, we have $\tau_{\rm rev} \to \infty$,
implying that if Onsager reciprocity holds, then the reversible 
operation is obtained {\it only} in the quasi-static limit. In other words, 
Onsager reciprocity must be violated for the heat cycle 
undergoing reversible operation in a finite duration.

The power output at reversible operation, 
$P_{\rm rev} = W_{\rm rev}^{(\tau)} / \tau_{\rm rev}$,
is given by:
\be
P_{\rm rev} = \frac{L_{12}^{2} - L_{21}^{2}}{4 L_{11}}\frac{\eta_{\rm C}^{2}}{T_c}.
\label{prev}
\ee
Thus, for a given value of $L_{12}$, 
the power output at reversible operation vanishes at both the extreme
values of $L_{21}$ (Eq. (\ref{1221})); at its minimum value, we have zero work output,
whereas at the maximum value of $L_{21}$, the Onsager reciprocity holds, leading to a diverging 
duration of the reversible cycle.

Using the second inequality in Eq. (\ref{1221}), we can write 
$(L_{12}/L_{11}) \eta_{\rm C}^{} \geq  
(L_{12} + L_{21}) \eta_{\rm C}^{} / 2 L_{11}$, or 
$ W_{\rm qs} \geq W_{\rm rev}^{(\tau)}$, 
where the equality is obtained when $\tau_{\rm rev} \to \infty$ (see Eq. (\ref{taurev})). 
Thus, we have the interesting result that the finite-time reversible
work is bounded from above by the quasi-static
work. 
Based on the results obtained so far, we may order the different work 
outputs as follows:
\be 
W_{\rm rev}^{(\tau)} \leq W_{\rm qs} \leq   W_{\rm rev}.
\label{3w}
\ee
Fig. 2 clarifies the meaning of the above inequalities 
in terms of relative magnitudes of the Onsager coefficients.
Clearly, the equalities are obtained  when
Onsager reciprocity is satisfied and the reversible
work is obtained just in the quasi-static limit.

\begin{figure}[ht]
 \includegraphics[width=8.5cm]{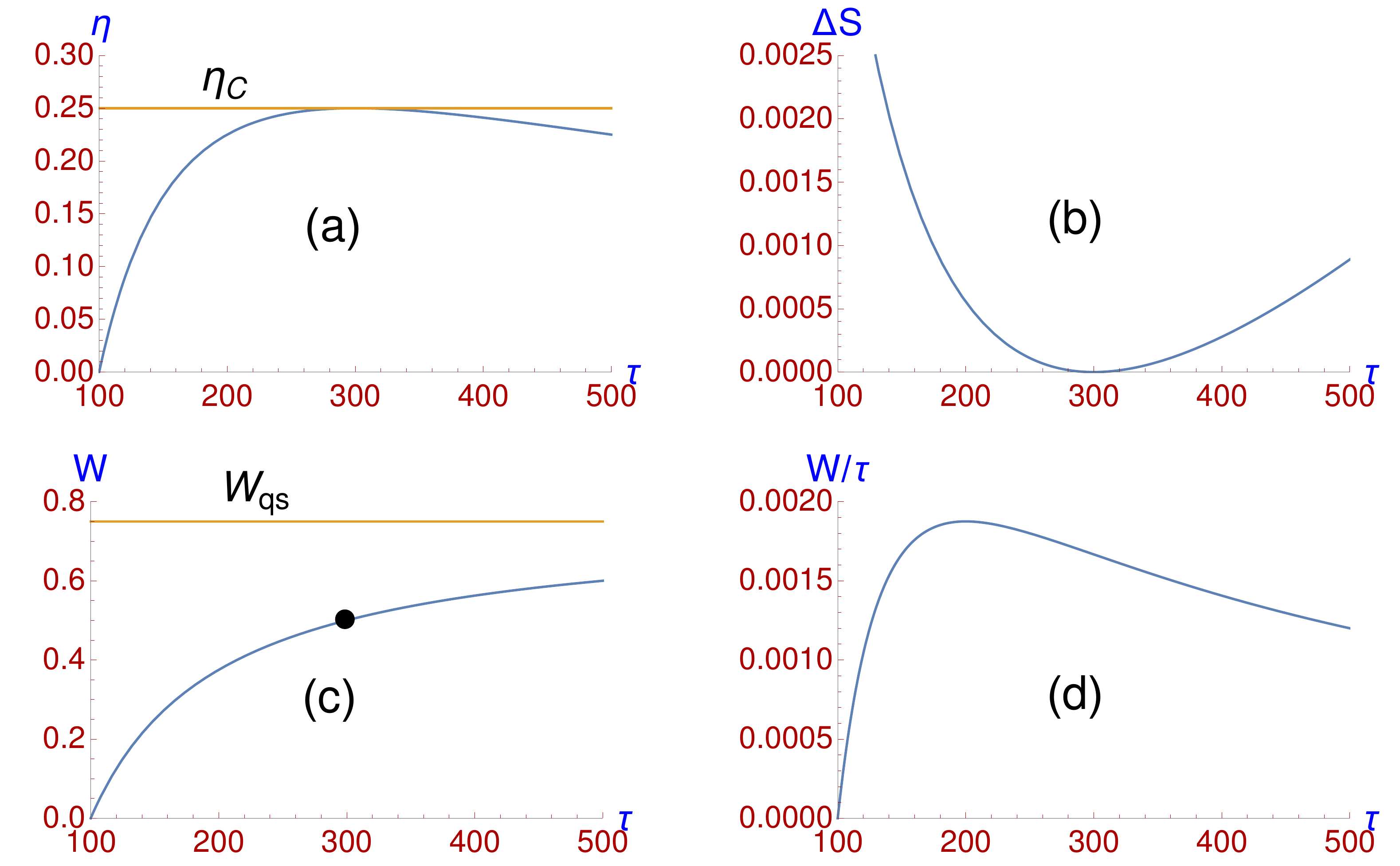}
\caption{An example of the $R^{(\tau)}_{}$ heat engine, 
showing reversible operation in a cycle of finite duration.
 The parameters are set at $L_{11} = 1, L_{22} = 4, L_{12} = 3$ and  
 $L_{21} =1$, consistent with the conditions $L_{12} >0$,  Eqs. (\ref{3rd}) and (\ref{1221}). 
 Also, $T_c = 75$ and $T_h = 100$, so that $\eta_{\rm C}^{} = 0.25$. 
 All quantities are plotted versus the cycle duration $\tau$ whose minimum value is 
 $\tau_{\rm min} = T_c/L_{12}\eta_{\rm C}^{} = 100$ units.
 (a) Efficiency $\eta$ obtains the Carnot bound at duration $\tau_{\rm rev} =
 300$ (see Eq. (\ref{taurev})). 
 For longer durations, the efficiency vanishes since $Q_h$ diverges.
 (b) Total increase in entropy per cycle, $\Delta S$, vanishes at $\tau_{\rm rev}$. 
 (c) Work output approaches the quasi-static limit $W_{\rm qs}$ 
 for long cycle durations. The work at reversible operation $W_{\rm rev}^{(\tau)}$
 ($\bullet$) is lower than $W_{\rm qs}$.
    (d) Power output becomes maximum at $\tau^* = 2 \tau_{\rm min} = 200$.
    The corresponding EMP is given by Eq. (\ref{emprev}). $\bullet$ denotes
    the power output under reversible conditions (Eq. (\ref{prev})).
 }
\end{figure}

\begin{figure}[ht]
 \includegraphics[width=9cm]{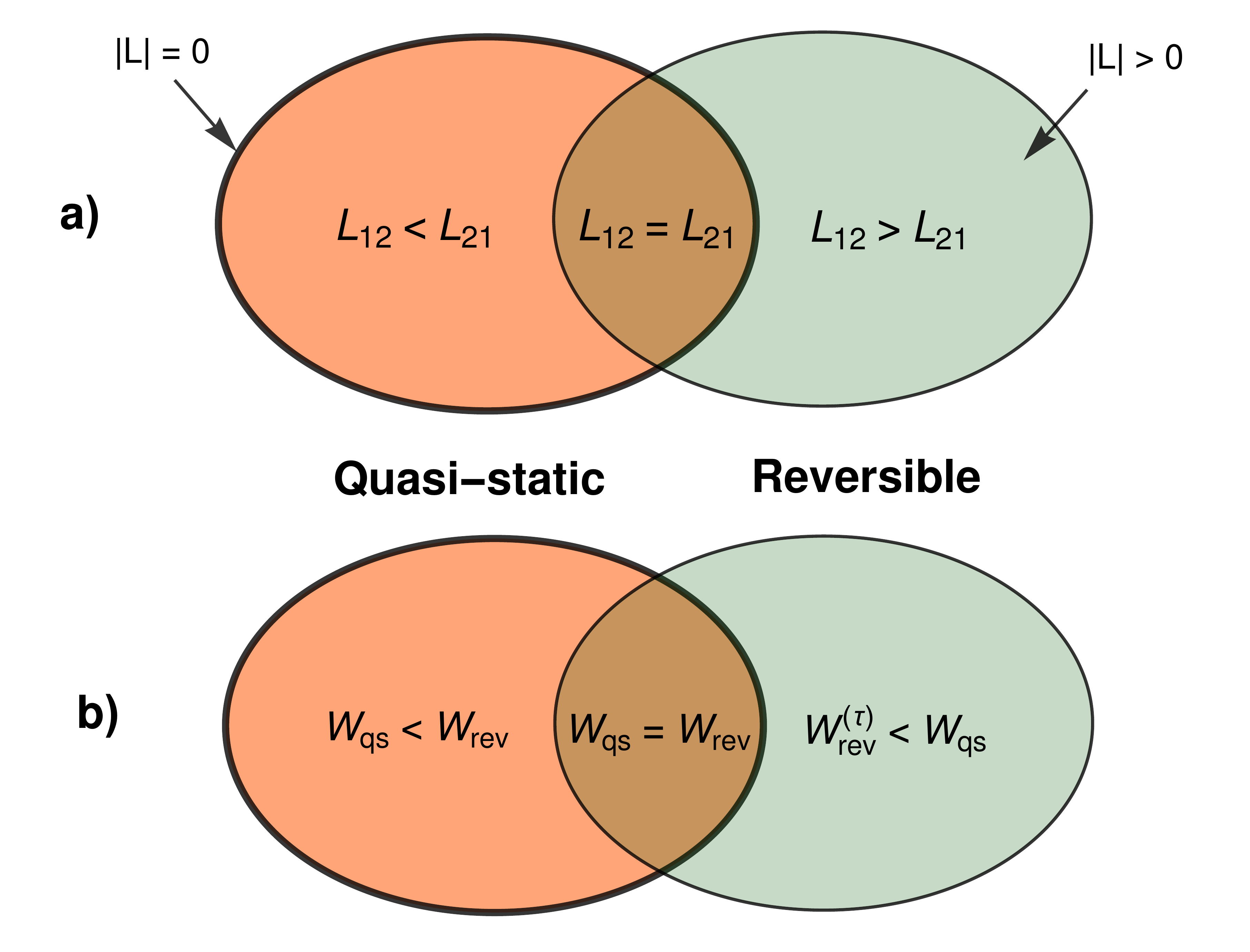}
 \caption{a) A cyclic, linear-irreversible engine 
  follows the Onsager reciprocal relation ($L_{12} = L_{21}$)
 if it approaches the quasi-static and the reversible operations, simultaneously.
 We have $L_{12} < L_{21}$, when the quasi-static cycle is not reversible,
whereas $L_{12} > L_{21}$ holds when the reversible cycle  
is not quasi-static, or in other words, achieved in a finite duration.
b) The corresponding comparison between different magnitudes of 
work output per cycle, also illustrating the inequalities in Eq. (\ref{3w}).
 }
\end{figure}

Exhibiting a reversible cycle in a finite duration for 
the linear-irreversible heat engine
is our second main result. For convenience, we denote 
such a heat engine as $R^{(\tau)}_{}$, which requires 
the conditions $ \vert {\bf L} \vert > 0$ and $L_{12}>0$, apart
from Eqs. (\ref{3rd}) and (\ref{1221}), to be satisfied.
Fig. 1 shows a case study of such an engine, where 
the Onsager coefficients can be chosen so as to 
achieve reversible cycle with finite duration.
Further, it can be easily shown that $\vert {\bf L} \vert = 0$ case
does not allow reversible cycle in a finite duration, since
it would violate the second law.

Finally, we study 
the optimization of average power output, $\dot{W} = W/\tau$. 
Setting $d\dot{W}/d\tau =0$, the optimal duration of the cycle 
is: $\tau^* =  2 \tau_{\rm min}$.
We note that the maximum power condition 
is not affected by whether $\vert {\bf L} \vert > 0$ or 
$\vert {\bf L} \vert = 0$. The work performed
at maximum power is $W^* = W_{\rm qs}/2$. So, 
the maximum power is 
$\dot{W}^* = W^*/\tau^* =  L_{12}^{2}\eta_{\rm C}^{2}/4L_{11}T_c$, 
and the efficiency at maximum power (EMP), $\eta^* = W^*/Q_h$, is:
\be
\eta^* = \frac{L_{12}^{2}}{2 \vert{\bf L}\vert + L_{12} L_{21}}
\frac{\eta_{\rm C}^{}}{2}.
\label{emprev}
\ee
The above expression was also derived as EMP for an autonomous engine
based on a thermoelectric setup in the presence of external magnetic field \cite{Benenti2011} 
and can be expressed in terms of two parameters (apart from $\eta_{\rm C}^{}$): 
the asymmetry ratio ($x\equiv L_{12}/L_{21}$)
and the generalized figure of merit ($y\equiv L_{12} L_{21}/ \vert{\bf L}\vert$). 
For $ \vert {\bf L} \vert = 0$ or the tight-coupling condition, 
the above formula is simplified to
$\eta^* = {\eta_{\rm qs}^{}}/{2}$. 
Since $L_{12} \leq L_{21}$ for the tightly coupled case,
so we conclude that $\eta^* \leq \eta_{\rm C}^{}/2$. 
The upper bound of half-Carnot value is obtained when quasi-static and reversible operations 
are equivalent and so the 
reciprocal relation holds. 
This result was known for autonomous, 
linear-irreversible  engines obeying  
the tight-coupling condition along with Onsager reciprocity 
\cite{Caplan1965, Broeck2005}. 
It was further noted in Ref. \cite{Benenti2011} that there are no 
thermodynamic constraints on the allowed values of the real parameter $x$.
However, for the case of an $R^{(\tau)}$ engine, there is a
constraint on $x$ due to Eq. (\ref{1221}), so that we have
$-1 \leq 1/x \leq +1$, or $\vert x\vert \geq 1$.
Thus, for such engines, the EMP is simplified to the form:
\be
\eta^* = \frac{\eta_{\rm C}^{}}{1 + (1/x)^2}.
 \label{emprev}
 \ee 
Notably, the half-Carnot value, established earlier as the upper bound 
for EMP, is breached here. Moreover, EMP is a function only of the 
parameter $x$, and as  $\vert x\vert \to \infty$, the EMP can 
approach the Carnot bound.  

Concluding, we have considered the performance of a cyclic heat engine
 within linear-irreversible framework. It is remarkable 
  that the idealized processes of
 equilibrium thermodynamics have a bearing on the reciprocal properties
 of phenomenological coefficients describing the strength
 of couplings in the near-equilibrium regime. 
 Our main general conclusions are: 
Onsager reciprocal relation implies that the reversible operation
is obtained in infinite time or quasi-static limit.
Since $W_{\rm qs}$ is finite, so in order to obtain Carnot 
efficiency, $Q_h$ should also be finite in that limit.
This necessarily implies $ \vert {\bf L} \vert = 0$.
On the other hand, even for $ \vert {\bf L} \vert = 0$, we may 
have a quasi-static operation which is irreversible. This
requires violation of Onsager reciprocity---in particular, $L_{12} < L_{21}$.
When $ \vert {\bf L} \vert > 0$, $Q_h$ diverges in the quasi-static limit
and so the efficiency vanishes. Interestingly, there is possibility of reversible
operation in a finite duration, for which  $L_{12} > L_{21}$ must be obeyed  
apart from other conditions mentioned in the text.
Thus, violation of the Onsager reciprocal relation implies that
quasi-static and reversible operations are not equivalent  
(see also Fig. 2). It may be remarked that our treatment
of Onsager reciprocity and its violation has been more at 
an abstract level, without going into the aspect of  
concrete, physical realizations. Even so, the consistency of some of 
the results with the steady-state thermoelectric setups \cite{Benenti2011}, 
such as the constraints for reversible operation in 
finite duration, 
indicate a wider basis for these results. 

The model considered here involves only two forces, which 
is the simplest example of an irreversible system exhibiting coupled processes.
Several lines of inquiry may be visualized in this context.
It would be interesting to highlight the parallels relating 
Onsager reciprocity with vanishing fluxes and reversible operations
in the case of autonomous engines.
A generalization incorporating more than two forces and/or
larger number of heat reservoirs is important.  
Apart from power output, the study of other figures of merit including 
 refrigerator models will be desirable. 
 Finally, the study of implications for stochastic thermal machines 
would foster an interesting line of inquiry.


\begin{thebibliography}{15}
\bibitem{commentqsrev} A quasi-static operation and a reversible operation
  are said to be equivalent if one implies the other.
\bibitem{commentcycles} A finite-time, irreversible Carnot cycle is 
expected to approach
the ideal or reversible Carnot cycle in the quasi-static limit,
whereas a finite-time Otto cycle should approach the quasi-static Otto 
cycle which is not reversible.
\bibitem{commentleaks}
$Q_h$ may diverge as $\tau\to \infty$ such that $J_2 = Q_h/\tau$ 
remains finite. In this limit, 
$J_2 = (\vert {\bf L} \vert /L_{11}) \eta_{\rm C}^{}/T_c$ with 
$\vert {\bf L} \vert/L_{11}$ as heat conductivity. 
This may happen if the heat leaks cannot be neglected.
\end{thebibliography}
\end{document}